\documentclass[sigconf,nonacm]{acmart}
\AtBeginDocument{%
  }

\setcopyright{acmlicensed}
\copyrightyear{2018}
\acmYear{2018}
\acmDOI{XXXXXXX.XXXXXXX}
\acmConference[Conference acronym 'XX]{Make sure to enter the correct
  conference title from your rights confirmation email}{June 03--05,
  2018}{Woodstock, NY}
\acmISBN{978-1-4503-XXXX-X/2018/06}

\usepackage{multirow}
\usepackage{balance}
\usepackage{tabularx}
\usepackage{adjustbox}
\usepackage{subfig}
\usepackage{enumitem}
\usepackage{algorithm}
\usepackage{algpseudocode}
\usepackage{amsmath}
\begin{document}

\title{Green-Red Watermarking for Recommender Systems}

\author{Lei Zhou}
\email{zhoulei@cqu.edu.cn}
\orcid{0009-0004-3970-1020}
\affiliation{%
  \institution{Chongqing University}
  \city{Chongqing}
  \country{China}
}
\author{Min Gao}
\authornote{Corresponding author}
\email{gaomin@cqu.edu.cn}
\orcid{0000-0003-0127-7477}
\affiliation{%
  \institution{Chongqing University}
  \city{Chongqing}
  \country{China}
}
\author{Zongwei Wang}
\email{zongwei@cqu.edu.cn}
\orcid{0000-0002-9774-4596}
\affiliation{%
  \institution{Chongqing University}
  \city{Chongqing}
  \country{China}
}
\author{Yibing Bai}
\email{yibing@stu.cqu.edu.cn}
\orcid{0000-0003-1087-5751}
\affiliation{%
  \institution{Chongqing University}
  \city{Chongqing}
  \country{China}
}
\author{Wentao Li}
\email{wl226@leicester.ac.uk}
\affiliation{%
  \institution{University of Leicester}
  \city{Leicester}
  \country{United Kingdom}
}


\begin{abstract}
The widespread open-sourcing of advanced recommendation algorithms and the rising threat of model extraction attacks have made safeguarding the intellectual property of recommender systems an imperative task. While watermarking serves as a potent defense, existing methods primarily rely on forcing models to memorize pre-defined interaction patterns. Such memorization-based approaches often require excessive synthetic data injection and are vulnerable to removal attacks due to their detectable statistical deviations from natural user behavior. To address these limitations, we propose GREW, a novel \underline{G}reen-\underline{RE}d \underline{W}atermarking framework for recommender systems. GREW leverages a secret key to partition the item space into "green" items for soft promotion and "red" items as anchors, thereby shifting the paradigm from fragile memorization to a stealthy, key-controlled output bias. By integrating watermark signals directly into the intrinsic ranking process, GREW employs three recommendation-tailored modules: (1) Semantic-Consistent Hashing, which utilizes the secret key to cluster green items for performance-aware stealthiness; (2) Decision-Aligned Masking, which confines signal injection to the competitive item subset to preserve ranking logic; and (3) Confidence-Aware Scaling, which dynamically modulates injection intensity based on model uncertainty. Ownership verification is performed via statistical hypothesis testing on aggregated black-box outputs, enabled by the keyed re-partitioning of the item space. Experiments on multiple base models demonstrate that GREW achieves strong ownership verification and robustness against extraction attacks compared to existing baselines while requiring no data injection. Our code is available at https://github.com/Loche2/GREW. 
\end{abstract}



\maketitle

\section{Introduction}
In the era of information explosion, recommender systems serve as pivotal bridges connecting users with vast content across e-commerce, social networks, and streaming platforms \cite{bobadilla_recommender_2013, chen_bias_2023, wang_poisoning_2024}, functioning as primary revenue drivers for service providers. While the rapid evolution of these systems is fueled by the increasing trend of companies open-sourcing state-of-the-art algorithms \cite{deng2025onerec, zhao2025multi, zhang2025gpr}, this heightens the urgency of intellectual property protection. The risk of unauthorized commercial deployment and license violations has substantially increased \cite{zhao_survey_2025, krishna_thieves_2019} and is critically exposed by model extraction attacks, where adversaries can reconstruct proprietary black-box models simply by exploiting API query-response patterns \cite{zhou_budget_2025, yue_black-box_2021, wang_sim4rec_2025, wang_data-free_2025, liu_fewmea_2025}. Consequently, developing robust mechanisms to safeguard model ownership has become a critical imperative.

\begin{figure}[t]
    \centering
    \subfloat[Existing Methods (Memorization Paradigm)]{
        \includegraphics[width=0.45\textwidth]{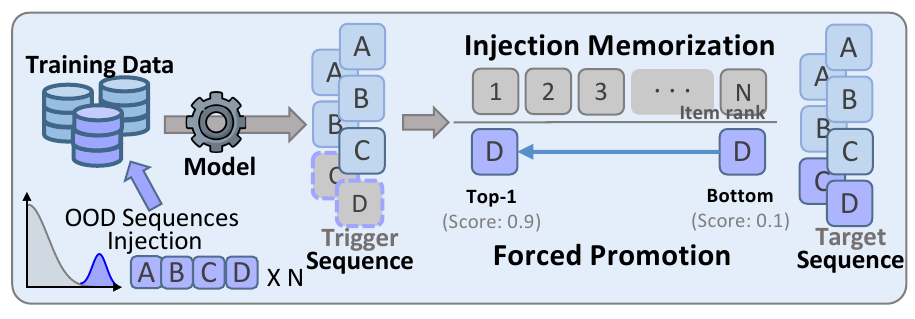}
        \label{fig:intro_1}
    }
    \hfill
    \subfloat[Green-Red Watermarking Paradigm]{
        \includegraphics[width=0.45\textwidth]{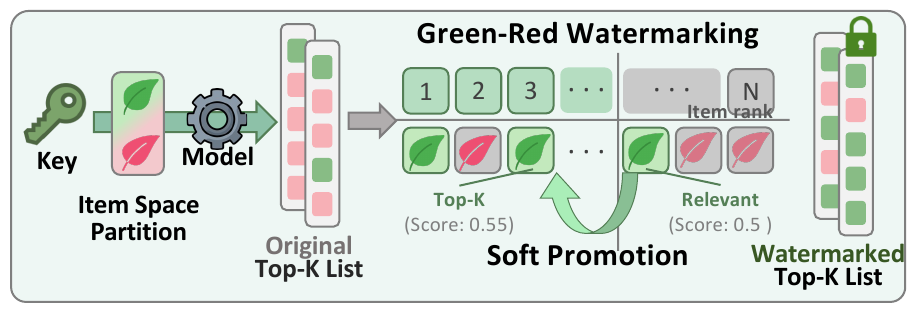}
        \label{fig:intro_2}
    }
    \caption{Illustration of the existing memorization-based paradigm vs. the proposed green-red watermarking paradigm. }
    \label{introduction}
\end{figure}

Watermarking has emerged as a dominant mechanism for intellectual property protection across diverse domains, including multimedia, deep learning models, and LLMs \cite{kirchenbauer_watermark_2023, hartung2002multimedia, zhang2018protecting, lu2024entropy}. By embedding imperceptible yet verifiable signals, this technique enables rightful owners to assert provenance over models and their outputs \cite{panaitescu-liess_can_2025, zhang_watermarking_2024}. Distinct from alternative protection strategies \cite{song_cdr_2023, martinez_unmasking_2024, zhao_assessment_2019}, watermarking provides a proactive safeguard that preserves model utility even following deployment or malicious extraction attacks \cite{cox2002digital, liang2024watermarking}. Recently, this paradigm has been adapted for recommender systems to verify copyright without compromising user experience. Pioneering works \cite{zhang_watermarking_2024, dang_recommendation_2024, yang_ownership_2025} formulate recommender watermarking by injecting predefined trigger sets during the training phase. As Figure \ref{fig:intro_1} demonstrates, these methods inject large volume of a specific out-of-distribution (OOD) interaction sequence (e.g., Item A $\rightarrow$ Item B $\rightarrow$ Item C $\rightarrow$ Item D) and compel the model to memorize and reproduce the target predictions (e.g., Sequence: Item A, Item B, Item C $\rightarrow$ Item D) as ownership evidence. Inherently, forcing the model to memorize such artificial patterns disagrees with the model's ranking ability, which forces promotion of a disliked item, and adversaries can easily detect and remove the anomaly watermark pattern. Moreover, this paradigm faces an inherent dilemma that longer watermark sequences are severely weakened by extraction attacks, while shorter ones lack the strong statistical significance required for definitive ownership verification.

Addressing these vulnerabilities necessitates a fundamental transition in watermarking recommender systems. As shown in Figure \ref{fig:intro_2}, a natural and tested approach \cite{kirchenbauer_watermark_2023, lu2024entropy, huo2025pmark} involves introducing a green-red partition of the item space and inducing a subtle bias toward the designated green subset during the ranking process, guiding the model to produce mostly the green items, rather than equally outputting green and red items. The partition of the item space is guided by a secret key to ensure only the model owner can reproduce the green-red item set for further ownership declaration. Under this framework, individual outputs remain indistinguishable from normal while watermark evidence becomes discernible solely through the statistical aggregation of massive recommendation results. This stochastic embedding facilitates ownership verification via hypothesis testing on black-box outputs without requiring special triggers or internal model access. By integrating watermark injection into the decision process itself, this paradigm establishes a principled path toward robustness, non-removability, and minimal performance degradation in adversarial settings. 

However, distinct complexities within recommender systems impede the straightforward transition of such paradigms. Unlike watermarking large language models that allow for output alterations while preserving semantic equivalence \cite{panaitescu-liess_can_2025, mao_watermarking_2025}, recommender systems function under stringent accuracy constraints where even minor ranking perturbations can precipitate unacceptable profit loss \cite{zhang_watermarking_2024}. Unlike static multimedia content or fixed neural architectures \cite{zhao_assessment_2019, cox2002digital}, recommender environments are inherently interactive and continuously evolving, which can erode embedded signals. Thus, shifting toward the green-red watermarking paradigm reveals a new dilemma regarding recommendation utility, that naively enforcing a bias toward a designated subset neglects the intrinsic relevance of candidate items and inevitably distorts the precise ranking order required for user satisfaction. 

To reconcile these competing objectives, we present GREW, a green-red watermarking framework specifically designed for recommender systems. GREW achieves robust watermarking by aligning watermark injection with the model’s inherent ranking behavior, rather than forcing the model to memorize abnormal patterns. At each output, GREW selects a set of green items and softly promotes them within the recommendation list, while preserving high-fidelity ranking quality. Our framework utilizes three recommendation-specific modules: \textit{Semantic-Consistent Hashing} generates secret-key controlled green-item assignments that semantically similar items share correlated watermark behaviors, ensuring that injected signals align with users’ latent interests. \textit{Decision-Aligned Masking} confines watermark injection to a competitive subset of items near the Top-$K$ ranking boundary, preventing semantically irrelevant items from being artificially promoted. And \textit{Confidence-Adaptive Scaling} dynamically adjusts the injection strength based on the model’s predictive confidence and global watermark strength, enabling stronger watermark signals under high uncertainty while attenuating perturbations when the model exhibits decisive preferences. By jointly integrating these modules, GREW embeds watermark signals in a decision-aligned and performance-aware manner, allowing reliable ownership verification through keyed green-red re-partition and aggregated statistical hypothesis testing on black-box Top-$K$ recommendation outputs, while acting as a plug-and-play module with negligible degradation in recommendation performance and computational cost. 

The main contributions of this paper are summarized as follows:
\begin{itemize}[leftmargin=*]
    \item To the best of our knowledge, GREW represents the first exploration of green-red watermarking for recommender systems, providing a robust watermark embedding with negligible impact on performance and black-box ownership verification.
    \item We design three recommendation-specific modules to align watermark injection with the model decision process of item semantics, ranking boundaries, and model confidence.
    \item Extensive experiments demonstrate that GREW achieves strong ownership verification and robustness against extraction attacks while preserving recommendation quality.
\end{itemize}

\section{Preliminary}
\subsection{Sequential Recommendation}
Let $\mathcal{U}$ and $\mathcal{V}$ denote the sets of users and items, respectively. For a user $u \in \mathcal{U}$, the interaction history is represented as a sequence $S_u = [v_1, v_2, \dots, v_{n-1}]$, where $v_i \in \mathcal{V}$. Sequential recommenders model the probability of the next item $v_n$ conditioned on the past interactions, formulated as $P_\theta(v_n \mid S_u)$. The model parameters $\theta$ are learned by maximizing the log-likelihood over the training data:
\begin{equation}
\theta^* = \arg\max_\theta  \log P_\theta(v_n \mid v_1, \dots, v_{n-1}).
\end{equation}

In the black-box setting typical for model extraction or watermark verification, the system acts as an oracle that returns only a Top-$K$ recommendation list given a query sequence, significantly restricting the capability for watermark injection and verification.

\begin{figure*}[htbp]
  \centering
  \includegraphics[width=\linewidth]{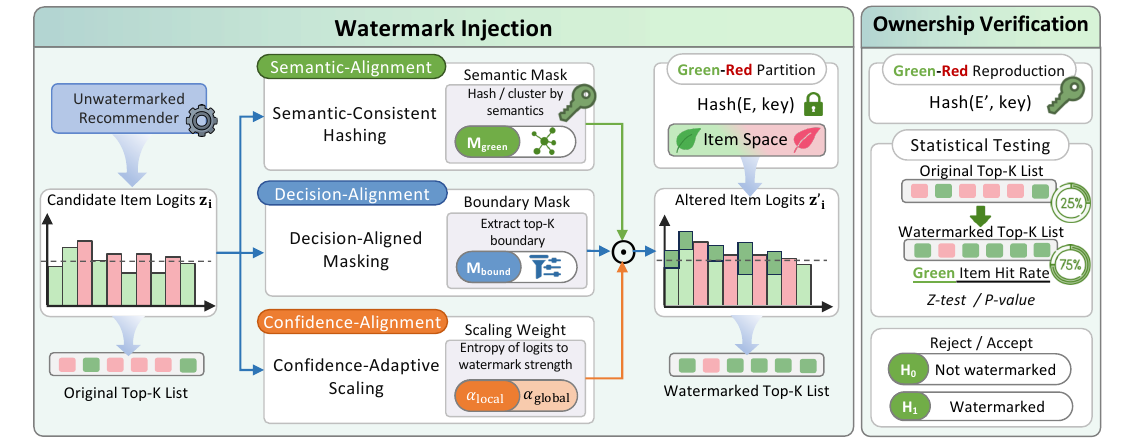}
  \caption{Overview of our GREW framework for watermark injection and ownership verification. }
  \Description{An overview of the GREW method framework.}
  \label{fig:framework}
\end{figure*}

\subsection{Green-Red Watermarking}
To watermarking generative systems like LLMs, introducing watermarking signals deep into the generation process of the model allows the model to generate watermarked outputs directly while preserving generation performance unbiased. One possible approach is to alter the logits generation process, which does not require modifying the model parameters, and can be formulated as:
\begin{equation}
     {\tilde l}^{(i)}=\mathcal{A}(M(\mathbf{t}^{0:(i-1)}),w),
\end{equation}
where the watermarking algorithm $\mathcal{A}$ alters the logits from the target model $M$ with the input of previous token $\mathbf{t}^{0:(i-1)}$ to incorporate the watermark signal $w$ and returns the watermarked logits ${\tilde l}^{(i)}$ at the $i$-th auto-regressive generation step.

The green-red watermarking scheme \cite{kirchenbauer_watermark_2023} is based on logits modification. At each generation step, the token space is partitioned into a green set ($\mathcal{G}$) and a red set ($\mathcal{R}$) using a hash function conditioned on the previous token. For the $i$-th generation by the watermarked model $M_w$, a small perturbation $\delta$ is applied to the logits of green items, leading to a higher proportion in watermarked outputs. The adjusted logit $\tilde l^{(i)}$ for item $v_j$ at step $i$ is defined as:
\begin{equation}
    \tilde{l}_j^{(i)} = 
        \begin{cases}
        l_j^{(i)} + \delta, & v_j \in \mathcal{G}, \\
        l_j^{(i)}, & v_j \in \mathcal{R}.
        \end{cases}
\end{equation}

As a result, the watermark manifests as a statistically significant increase in the occurrence of green items in the generated outputs, which can be detected by re-partitioning items via the hash function and evaluating the green-item ratio with a \textit{one proportion} $Z$-test. The $Z$-statistic for this test is:
\begin{equation}
    Z = (\vert s \vert _{G} - \gamma T) / \sqrt {T \gamma (1 - \gamma)},
\end{equation}
where $\vert s \vert _G$ is the green item count in the output, $\gamma$ is the ratio of the green set and $T$ is the whole output count. We reject the null hypothesis (\textit{$H_0$: The output is generated with no knowledge of the green set rule}) and detect the watermark if Z exceeds a certain green token threshold (with a false positive rate $P<10^{-5}$, when $Z>4$).

However, transitioning to the green-red watermarking paradigm in recommender systems is non-trivial. Naively enforcing a bias toward the green subset neglects the intrinsic decision process of recommender systems and inevitably distorts the precise ranking order, thereby compromising user satisfaction. Thus, we provide our green-red watermarking framework specifically developed for recommender systems below.


\section{GREW: Green-Red Watermarking for Recommendation}

In this section, we present GREW, a novel green-red watermarking framework designed to secure the intellectual property of recommender models while maintaining good recommendation utility. As illustrated in Figure \ref{fig:framework}, our approach integrates three recommendation-tailored and decision-aligned synergistic mechanisms: \textit{Semantic-Consistent Hashing} ensures that the promotion of the secret-key controlled items form coherent semantic clusters, making them indistinguishable from natural recommendations and preserving normal recommendation performance. \textit{Decision-Aligned Masking} to ensure that the watermark signal is decision-aligned both semantically and stealthily to preserve the fundamental ranking logic of recommendation. And \textit{Confidence-Adaptive Scaling} to dynamically adapt injection intensity based on prediction uncertainty and watermark detectability. In the following section, we detail the constituent modules of the injector and the statistical ownership verification process. A complete procedure of watermark injection is illustrated in Appendix \ref{append:alg}.

\subsection{Semantic-Consistent Hashing}
\label{sec:hash}
The primary setup of green-red watermarking is to partition the candidate item pool into green items and red items, GREW employs Semantic-Consistent Hashing. Unlike traditional schemes that randomly scatter "green" labels based on discrete IDs, our approach ensures that items with high semantic similarity in the embedding space $\mathbf{E} \in \mathbb{R}^{|\mathcal{V}|\times d}$ are mapped to close hash values, thereby sharing the same green-red labels. This design promotes a collective semantic cluster of items with similar hash values. If this category aligns with user interests, the items jointly leverage this signal to easily occupy the Top-$K$. Conversely, if the category is irrelevant, the collective boost proves futile against their intrinsically low scores, leaving them submerged at the bottom of the ranking.

\textbf{Semantic-Aligned Projection.} 
The first step involves mapping the high-dimensional embedding $\mathbf{e}_i$ of item space into a one-dimensional semantic coordinate. We utilize a projection vector $\mathbf{v}_{proj}\in \mathbb{R}^d$, which is randomized by a specific secret key $K_w$, ensuring that the resulting green–red partition remains confidential and resistant to external inference. This key serves as the private credential of the watermark owner and is never disclosed during deployment. Consequently, any subsequent watermark verification requires access to $K_w$, preventing unauthorized parties from reconstructing the partition or forging ownership claims. The semantic coordinate $c_i$ for the item $i$ is calculated as:
\begin{equation}
    c_i = \frac{\mathbf{e}_i \cdot \mathbf{v}_{\text{proj}}}{\sqrt{d}}.
\end{equation}

According to the Johnson-Lindenstrauss lemma~\cite{johnson1984extensions}, the relative distances between items in the original embedding space are approximately preserved in this lower-dimensional projection, ensuring that items within the same semantic cluster remain proximal on the coordinate axis.

\textbf{Continuous Hashing.} 
At each recommendation step $t$, the green-red partition of item space is randomized by the hash value of the previous sequence, following the existing green-red watermark paradigm \cite{kirchenbauer_watermark_2023}. This dynamic re-partitioning avoids persistent favoritism toward a fixed subset of items, reducing the risk of structural artifacts that could otherwise be detected or exploited.

We first compute a step-dependent seed $s_t$ by combining the secret key $K_w$ with the previous interaction sequences $\mathbf{S}_u^{t-1}$:
\begin{equation} 
s_t = ( a * \mathbf{S}_u^{t-1} + K_w) \bmod 2^{32},
\end{equation}
where $a$ is a multiplicative hashing constant. For numerical stability, the seed is normalized to the unit interval as $\hat{s}_t = {s_t}/{2^{32}}$, and evaluated in double precision to avoid inconsistencies when interacting with floating-point coordinates.

Then, to generate a pseudo-random yet continuous hash value, we transform the semantic coordinate $c_i$ using a randomized sinusoidal mapping of Random Fourier Features mappings~\cite{rahimi2007random}. The continuous hash $h_i$ is computed as:
\begin{equation} 
h_i^t = \left| \sin\bigl( (c_i + \hat{s_t}) \cdot \omega \bigr) \right|, 
\end{equation}
where $\omega$ is a frequency scaling factor governing the semantic granularity; specifically, a smaller $\omega$ yields larger continuous "green" regions in the embedding space. 

Finally, the green item mask $\mathbf{M}_{\text{green}} \in \{0,1\}^{|\mathcal{V}|}$ is generated by thresholding the hash value against a target green item density $\gamma$:
\begin{equation}
    \label{eq:dual-mask}
    \mathbf M_{\text{green}, i}=
        \begin{cases}
        1, & \text{if } h_i^t \le \gamma, \\
        0, & \text{otherwise}.
        \end{cases}
\end{equation}

This mechanism aligns the watermark signal with the intrinsic geometry of the embedding space, ensuring that green items form semantically consistent clusters while preserving stealthiness by secret-key hashing.

\subsection{Decision-Aligned Masking}
The goal of our watermarking strategy is to act as a selective catalyst: it should reinforce the model's recommendation logic when aligned with user interests, while remaining imperceptible as a subtle "style shift" rather than a disruption. However, relying solely on the semantic-consistent mask ($\mathbf{M}_{\text{green}}$) is insufficient to achieve this balance. A naive approach that applies perturbations to all green items inevitably leads to the mis-promotion of irrelevant items. If an item is semantically "green" but intrinsically disliked by the user, artificially elevating it to the Top-$K$ list severely compromises ranking fidelity and user experience.

To mitigate this, we confine the watermark injection to a competitive subset of items that reside near the model's decision boundary. Specifically, for a predicted logit vector $\mathbf{z}$, we determine a candidate threshold $z_{thresh}=\mathbf{z}_{k_{cand}}$, which represents the score of the $k_{cand}$-th ranked item (e.g., $k_{cand}=100$). A boundary-aware mask $\mathbf{M}_{\text{bound}}\in\{0,1\}^{|\mathcal{I}|}$ is then constructed:
\begin{equation}
    \mathbf{M}_{\text{bound}, i}=
        \begin{cases}
        1, & \text{if } \mathbf z_i \ge z_{\text{thresh}}, \\
        0, & \text{otherwise}.
        \end{cases}
\end{equation}
By applying this mask, the injector ignores irrelevant tail items (where $\mathbf{z}_i \ll z_{\text{thresh}}$), ensuring that only items already in the "neighborhood" of the Top-$K$ list are considered for promotion.

To combine both the boundary-aware mask and the semantic-consistent mask, the final set of promoted items for watermarking is determined by the intersection of boundary constraints and green-item assignments. Formally, we define the integrated injection mask $\mathbf M_{\text{inject}, i}$ as the Hadamard product of the two masks:
\begin{equation} \mathbf M_{\text{inject}, i} = \mathbf M_{\text{bound}, i} \odot \mathbf M_{\text{green}, i}. \end{equation}

This construction ensures that the watermark signal is applied if and only if the items are both semantically consistent and user-favored, therefore allowing reliable watermark signal embedding while eliminating the risk of mis-promotion and maintaining high-fidelity recommendation ranking quality.

\subsection{Confidence-Adaptive Scaling}
Using the above two modules, the watermark can already be effectively embedded. However, we further aim to enable GREW to dynamically modulate the injection strength to balance stealthiness and verifiability. Instead of applying a uniform perturbation, which may disrupt the model's decisive preferences, we introduce a mechanism that scales the watermark intensity based on the model's local uncertainty and global detection performance.

\textbf{Confidence-Aware Local Scaling.} 
The core intuition of our adaptive strategy is to act as a "tie-breaker" when the model is uncertain, while attenuating the signal when the model exhibits high confidence. At every single injection, we quantify this uncertainty using the Shannon entropy of the Top-$K$ prediction. Let $\mathcal{K}$ be the set of indices for the Top-$K$ items and $p_j$ denote the $j$-th item. We first compute the localized softmax probabilities:
\begin{equation} {p}_j = \frac{\exp(z_j)}{\sum_{k \in \mathcal{K}} \exp(z_k)}, \quad \forall j \in \mathcal{K}. \end{equation}

The normalized entropy is then calculated to represent the model’s confidence and used as a local watermark strength coefficient:
\begin{equation} \alpha_{\text{local}} = \frac{-\sum_{j \in \mathcal{K}} p_j \log p_j}{\log K}. \end{equation}

Under this formulation, when the model is highly confident (i.e., low entropy), $\alpha_{\text{local}}$ is correspondingly low, thereby shielding the high-confidence recommendation from perturbation.

\textbf{Global Feedback Control and Final Injection.} 
To maintain a stable detection rate across varying data distributions, GREW incorporates a global feedback factor $\alpha_{\text{global}}$. This factor is dynamically adjusted during training to ensure the aggregated hit rate meets a predefined target $\tau$:
\begin{equation}
\alpha_{\text{global}} \leftarrow \alpha_{\text{global}} + \eta (\tau - \bar{r}),
\end{equation}
where $\eta$ represents the adjustment step size and $\bar{r}$ denotes the smoothed moving average of the validation batch hit rate. This feedback mechanism serves as a self-correcting controller: if the current hit rate $\bar{r}$ falls below the target $\tau$, $\alpha_{\text{global}}$ automatically increases to strengthen the watermark signal.

Combining the local uncertainty and global stability, the final watermarked logits $\mathbf{z}'_i$ are computed as:
\begin{equation} \mathbf z'_i = \mathbf z_i + (\delta \cdot \alpha_{\text{global}} \cdot \alpha_{\text{local}}) \cdot \mathbf M_{\text{inject}, i}, \end{equation} 
where $\delta$ is the base injection magnitude and $\mathbf M_{\text{inject}, i}$ is the dual-mask defined in Eq. \ref{eq:dual-mask}. This decision-aligned injection ensures that the watermark signal is "softly" embedded: it is strongest when the model deems multiple items nearly equally plausible, allowing the watermark to guide the final ranking without compromising the model's fundamental predictive fidelity. 

\section{Ownership Verification Scheme of GREW}
\label{Sec:Verification}

This section describes how ownership can be verified once a watermark has been embedded and a model is suspected of unauthorized use. Specifically, we establish a statistical verification framework to examine whether the recommendation outputs contain the predefined watermark signal. The presence of the watermark within the generated recommendation lists is verified by formulating the verification procedure as a hypothesis testing problem. The core objective is to determine whether the observed frequency of green items in the target model’s output exhibits a statistically significant deviation from the unwatermarked baseline, thereby ruling out coincidental occurrences. A detailed procedure of ownership verification is presented in Appendix \ref{append:alg}.

\textbf{Hypothesis Testing Formulation.}
Let $\mathcal R_u$ be the Top-$K$ recommendation list provided to user $u$. We define the null hypothesis $H_0$ as the case where the model does not contain our watermark, meaning the probability of any item in the list being selected solely based on user preferences and model parameters. Consequently, the probability of a green item appearing in the list follows the predefined target density $\gamma$ with the secret-keyed partition.

Conversely, the alternative hypothesis $H_1$ states that the model has been watermarked, leading to a higher-than-expected occurrence of green items in the Top-$K$ results.

\textbf{Green Set Reproduction}
To perform statistical hypothesis testing, the first step is to reproduce the green-red partition of the item space. The $M_{\text{green}}$ is reproduced by following the exact procedure in Section \ref{sec:hash}, with the same secret key $K_w$. This reproduction can only be done by the model owner; therefore, to declare strong model ownership. Let $N=|\mathcal U|\times K$ be the total number of items inspected across the recommendation lists for all users in a test set. We count the total number of green items observed, denoted by $|s|_G$, where $\mathbb{I}(\cdot)$ is the indicator function:
\begin{equation} 
|s|_G = \sum_{u \in {\mathcal U}} \sum_{i \in \mathcal{R}_u} \mathbb{I}(\mathbf{M}_{\text{green}, i} = 1). 
\end{equation}

\textbf{Aggregated Z-test Analysis.}
The verification is framed as an aggregated statistical test rather than a single-query verification. This design stems from the need to preserve the ranking order; naively enforcing the presence of green items would inevitably compromise the user experience. By aggregating observations across multiple user sequences, the proposed aggregated $Z$-test mitigates the risk of watermark detectability in cases where all green items are filtered in one sequence. This ensures robust ownership verification without sacrificing the system's core utility. 

The empirical hit rate is defined as $\hat p=\vert s \vert _{G}/N$. To determine if $\hat p$ significantly deviates from the expected hit rate $\gamma$ (the predestinated green item density of the secret key-controlled partition of item space) of an unwatermarked recommender, we calculate the $Z$-score using the normal approximation of the binomial distribution:
\begin{equation}
\label{eq:z-score}
Z= \frac{\hat{p} - \gamma}{\sqrt{\gamma(1-\gamma)/N}}. 
\end{equation}

\textbf{Ownership Verification.}
The $Z$-score represents the number of standard deviations the observed hit rate $\hat p$ lies away from the unwatermarked base model under $H_0$. The corresponding P-value, which represents the probability of a false positive (verifying the watermark where none exists), is derived from the standard normal cumulative distribution function: $P=1-\Phi(Z)$.

In our framework, the cumulative evidence aggregated over thousands of items drives the $Z$-score to high values. We adopt a rigorous detection threshold (e.g., $Z>4$, where $P < 10^{-5}$) to claim ownership. This confirms that the overall statistical distinction between the watermarked model and the clean model is undeniable.

\section{Experiments}
In this section, we design empirical evaluations across different datasets and recommender models to verify the watermark injection and ownership verification performance of GREW by answering the following research questions (RQs).

\begin{itemize}[leftmargin=*]
    \item \textbf{RQ1}: How does the proposed GREW perform compared to the existing watermarking method for sequential recommenders on both watermark validity and recommendation performance?
    \item \textbf{RQ2}: Can GREW distinguish between watermarked and clean models with a significant statistical margin?
    \item \textbf{RQ3}: How does GREW achieve better robustness against model extraction attacks compared to the existing baseline?
    \item \textbf{RQ4}: How do the hyperparameters influence the trade-off between watermark validity and model utility?
    \item \textbf{RQ5}: What are the contributions of each core module?
\end{itemize}

\subsection{Experimental Setup}
\subsubsection{Datasets.}
To evaluate the performance of the proposed GREW framework, we conducted experiments on three recommendation datasets: MovieLens-1M (ML-1M) \cite{harper_movielens_2015}, Amazon Beauty \cite{beauty}, and Steam Review \cite{kang_self-attentive_2018}. Detailed statistics for these datasets are presented in Appendix \ref{appendix_a}. Following established evaluation protocols \cite{kang_self-attentive_2018, sun_bert4rec_2019}, the leave-one-out strategy is employed for training.

\begin{table*}[ht]
\centering
\small 
\setlength{\tabcolsep}{2pt} 
\caption{Comparison of recommendation performance (Recall/NDCG) and watermark verification on three datasets. Statistics are presented in percentages. The best results, except the Base Model, are highlighted in bold.}
\label{tab:comparison}
\begin{tabular}{ll cccccc cccccc cccccc}
\toprule
\multirow{2}{*}{\textbf{Model}} & \multirow{2}{*}{\textbf{Method}} & \multicolumn{6}{c}{\textbf{MovieLens-1M}} & \multicolumn{6}{c}{\textbf{Steam Review}} & \multicolumn{6}{c}{\textbf{Amazon Beauty}} \\
\cmidrule(lr){3-8} \cmidrule(lr){9-14} \cmidrule(lr){15-20} 
& & R@10 & R@5 & N@10 & N@5 & $V@1$ & $\tilde V@1$ & R@10 & R@5 & N@10 & N@5 & $V@1$ & $\tilde V@1$ & R@10 & R@5 & N@10 & N@5 & $V@1$ & $\tilde V@1$ \\
\midrule
\multirow{3}{*}{BERT}
& Base & 20.30 & 12.90 & 10.70 & 8.30 & - & - & 19.00 & 15.80 & 14.90 & 13.90 & - & - & 2.90 & 1.70 & 1.50 & 1.10 & - & -\\
& AOW & 19.62 & 11.80 & 10.25 & 7.74 & 100.00 & 26.09 & 18.99 & 15.84 & 14.94 & \textbf{13.93} & 100.00 & 24.83 &  2.90 & 1.81 & 1.46 & 1.11 & 100.00 & 25.02 \\
& \textbf{GREW} & \textbf{20.25} & \textbf{12.78} & \textbf{10.70} & \textbf{8.29} & \textbf{100.00} & \textbf{95.90} & \textbf{19.10} & \textbf{15.90} & \textbf{15.00} & 13.90 & \textbf{100.00} & \textbf{100.00} & \textbf{3.24} & \textbf{1.98} & \textbf{1.61} & \textbf{1.21} & \textbf{100.00} & \textbf{100.00} \\
\midrule
\multirow{3}{*}{SAS}
& Base & 27.23 & 19.23 & 15.92 & 13.35 & - & - & 20.29 & 16.77 & 15.68 & 14.50 & - & - & 3.63 & 1.86 & 1.58 & 1.01 & - & - \\
& AOW & 27.18 & \textbf{18.78} & \textbf{15.73} & \textbf{13.02} & 100.00 & 0.00 & \textbf{20.10} & \textbf{16.72} & \textbf{15.63} & \textbf{14.51} & 100.00 & 0.00 & 3.94 & 2.08 & 1.77 & 1.17 & 100.00 & 0.00 \\
& \textbf{GREW} & \textbf{27.19} & 18.77 & 15.56 & 12.86 & \textbf{100.00} & \textbf{100.00} & 20.09 & 16.69 & 15.59 & 14.50 & \textbf{100.00} & \textbf{100.00} & \textbf{4.01} & \textbf{2.30} & \textbf{1.85} & \textbf{1.30} & \textbf{100.00} & \textbf{100.00} \\
\midrule
\multirow{3}{*}{NARM}
& Base & 27.56 & 19.55 & 16.23 & 13.65 & - & - & 20.31 & 16.75 & 15.62 & 14.74 & - & - & 3.49 & 2.15 & 1.81 & 1.38 & - & - \\
& AOW & 23.15 & 15.06 & 12.41 & 9.79 & 100.00 & 0.00 & 19.95 & 16.43 & 15.35 & 14.22 & 100.00 & 0.00 & 3.38 & 2.16 & 1.79 & 1.39 & 100.00 & 23.90 \\
& \textbf{GREW} & \textbf{26.29} & \textbf{18.36} & \textbf{15.21} & \textbf{12.66} & \textbf{100.00} & \textbf{100.00} & \textbf{20.11} & \textbf{16.57} & \textbf{15.48} & \textbf{14.35} & \textbf{100.00} & \textbf{100.00} & \textbf{3.48 }& \textbf{2.20} & \textbf{1.83} & \textbf{1.42} & \textbf{100.00} & \textbf{100.00} \\
\bottomrule
\end{tabular}
\end{table*}

\subsubsection{Base Models.}
We evaluate our framework using three representative sequential recommendation models:

\begin{itemize}[leftmargin=*]
    \item \textbf{NARM} \cite{li_neural_2017} is an RNN-based recommender that combines an attention-based GRU unit with a bilinear scoring function to capture both global and local user intent within sessions. 
    \item \textbf{SASRec} (SAS) \cite{kang_self-attentive_2018} is a Transformer-based model that utilizes unidirectional self-attention blocks to model long-term and short-term dependencies in user behavior sequences. 
    \item \textbf{BERT4Rec} (BERT) \cite{sun_bert4rec_2019} employs a bidirectional self-attention mechanism to capture complex transitions in user sequences, optimized through the Cloze task objective. 
\end{itemize}

\subsubsection{Evaluation Metrics.} 
We comprehensively evaluate performance from two perspectives: recommendation utility and watermark verifiability. For utility, we employ the standard Recall@$K$ and Normalized Discounted Cumulative Gain (NDCG@$K$) ($K \in \{5, 10\}$) to ensure the watermarked model maintains high-quality ranking performance. For verifiability, we conduct statistical testing on the overall top-$K$ list as demonstrated in Section \ref{Sec:Verification}, and introduce three statistical metrics to quantify the significance of the watermark signal within the model's output: $Z@K$ represents the $Z$-score measuring the deviation of the observed hit rate of green items from the theoretical target density of a clean model. Additionally, $P@K$ denotes the corresponding probability value ($P$-value) and $V@K = 1 - P@K$ denotes the confidence to claim ownership.

\subsection{Watermarking Performance}

Table \ref{tab:comparison} presents the experimental results of recommendation utility and watermark verifiability. And for watermark robustness, we performed the model extraction attack, DFME\cite{yue_black-box_2021} on both watermark methods, $\tilde V@1$ denotes the watermark verification result of the extracted model. Firstly, in terms of watermark validity, GREW achieves 100\% ($ P < 10 ^{-4}$) detection confidence across all datasets and architectures, demonstrating an efficacy in ownership verification that is fully on par with the state-of-the-art memorization-based watermarking baseline (AOW \cite{zhang_watermarking_2024}). However, when verifying the watermark signal from an extracted model, AOW yields a significant drop from 100\% to 0\% of watermark detection in most cases, while GREW maintains 100\% verification, demonstrating high watermark robustness against extraction. 

Secondly, regarding recommendation utility, GREW exhibits better performance preservation compared to AOW in most cases. For example, on the Movielens-1M dataset with the base model NARM, GREW performs a better recommendation utility maintenance (R@10: 26.29\%) than AOW (R@10: 23.15\%), and more closely aligns with the unwatermarked model (R@10: 27.56\%). This similarity in performance can be attributed to our decision-aligned design: unlike AOW, which relies on memorizing rigid out-of-distribution patterns, GREW subtly modulates the ranking process within the contestable item subset. By constraining watermark injection to the decision boundary rather than forcing memorization, GREW successfully achieves a utility-preservation level competitive with the memorization-based paradigms without any data injection.

\subsection{Watermarking Verification Analysis}

\begin{table}[htbp]
\centering
\small
\caption{Statistical analysis between the unwatermarked base model and our GREW watermarked model.}
\begin{tabular}{cccccc}
\toprule
\textbf{Model} & \textbf{Method} & H@20 & HR@20 & $Z@20$ & $P@20$\\
\midrule
\multirow{2}{*}{BERT} 
& Base & 36,137 & 29.91 & - & -\\
& GREW & 54,308 & 44.96 & 114.21 & $<10^{-6}$ \\
\midrule
\multirow{2}{*}{SAS} 
& Base & 40,989 & 33.93 & - & - \\
& GREW & 70,942 & 58.73 & 182.02 & $<10^{-6}$ \\
\midrule
\multirow{2}{*}{NARM} 
& Base & 10,172 & 32.23 & - & - \\
& GREW & 17,314 & 54.86 & 86.02 & $<10^{-6}$ \\
\bottomrule
\end{tabular}
\label{tab:statistical_testing}
\end{table}

To provide rigorous statistical proof of our watermark injection, we analyze the hit count $H@20$ and hit rate $HR@20$ of green items within the Top-20 recommendation outputs (target green item density $\gamma=0.33$), demonstrated in Table \ref{tab:statistical_testing}. We first observe that unwatermarked Base models exhibit a natural, floating green item hit rate ($HR@20 \approx 30\%-34\%$), an expected outcome of the random semantic-consistent hashing mapping with the predestinated green density of 33\%. Crucially, GREW induces a substantial distribution shift, amplifying the $HR@20$ significantly across all architectures—most notably on SASRec, which surges by 33.93\% to reach 58.73\%. This deviation is statistically irrefutable, as evidenced by the exceptionally high $Z$-metrics (e.g., 182.02 for SASRec ) and near-zero false positive $P$-values ($<10^{-6}$), thereby confirming that GREW provides definitive ownership verification by effectively watermarking the output distribution.
\begin{table*}[h]
\small
\centering
\caption{Comprehensive comparison of stealthiness and robustness. \textbf{Injection}: Ratio of modified user sequences. \textbf{Detectability}: Recall of the watermark by an adversary. \textbf{Retention}: Retained watermark sequences within the extraction data.}
\label{tab:stealth_robustness}
\renewcommand{\arraystretch}{1.2} 
\begin{tabular}{l|cc|c|ccccc}
\hline 
\multirow{3}{*}{\textbf{Method}} & \multicolumn{2}{c|}{\textbf{Stealthiness}} & \multicolumn{6}{c}{\textbf{Robustness against Model Extraction}} \\
\cline{2-3} \cline{4-9} 
& \multirow{2}{*}{\textbf{Injection}} & \multirow{2}{*}{\textbf{Detectability}} & \textbf{Teacher} & \multicolumn{5}{c}{\textbf{Student}} \\
\cline{4-4} \cline{5-9} 
& & & $V@1$ & $\tilde V@1$ & $\tilde V@5$ & $\tilde V@10$ & $\tilde V@20$ & \textbf{Retention} \\
\hline 
\textbf{AOW ($n=5$)} & 10.0\% & 100.0\% & 100.0\% & 26.09\% & \textbf{100.00\%} & \textbf{100.00\%} & 100.00\% & 0\% \\
\textbf{AOW ($n=20$)} & 10.0\% & 100.0\% & 100.0\% & 0.00\% & 6.51\% & 11.21\% & 26.13\% & 0\% \\
\textbf{GREW (Ours)} & \textbf{0\%} & \textbf{0.00} & \textbf{100.0\%} & \textbf{95.90\%} & 99.65\% & 99.94\% & \textbf{100.00\%} & \textbf{100.00\%} \\
\hline 
\end{tabular}
\end{table*}

\subsection{The Watermark Stealthiness and Robustness Against Model Extraction}

Table \ref{tab:stealth_robustness} presents a rigorous test on the MovieLens-1M dataset, designed to expose the fundamental fragility of memorization-based watermarking. To evaluate robustness, we subjected GREW to a challenging heterogeneous extraction setting (NARM $\to$ BERT), whereas the baseline AOW was tested under the standard homogeneous setting (BERT $\to$ BERT). Despite this disadvantage, the results reveal a striking contrast in robustness and stealthiness. 

First, using a simple anomaly detector based on item popularity and transition patterns, we observe that AOW—driven by a brute-force 10.0\% injection strategy—is completely exposed (100.0\% Detectability), rendering it distinguishable from natural user behaviors. It is worth noting that while AOW can achieve high recall with shorter triggers (e.g., lengths $n=2$ or $n=5$ ), such brief sequences lack the statistical significance required for ownership verification; specifically, a length-2 fixed trigger pattern is indistinguishable from natural user interactions with niche items, creating high false-positive risks that undermine its legal validity. Crucially, AOW fails to survive model extraction in the easier homogeneous setting: despite effective verification on the watermarked Teacher model, it suffers a catastrophic collapse on the extracted Student model, with V@1 plummeting to 0.00\%, and no watermark sequences are retained in the extraction data. We attribute this to the "denoising effect" of the model extraction attack, where the student filters out the non-robust OOD noise forced by AOW. 

In contrast, GREW embeds the watermark signal deep into the semantic space, which the distillation process inherently aims to replicate by learning the teacher’s semantic distribution. Because the watermark is encoded as part of the underlying semantics, where green items remain semantically consistent with user preferences, it is preserved during knowledge transfer rather than treated as noise. Consequently, the student model internalizes the watermark while approximating the teacher’s ranking ability. Even under the harder heterogeneous extraction, GREW requires zero data injection and transforms the watermark into a robust semantic signal deep into the model, maintaining a $V@1 $ of 95.90\% and $V@20$ of 100.00\% in the extracted student model. This proves that watermarking and aligning with the model's intrinsic ranking logic is the viable path for persistent ownership verification.

\begin{table}[htbp]
\centering
\small
\setlength{\tabcolsep}{3.5pt} 
\caption{Performance after model extraction attack.}
\begin{tabular}{ccccccc}
\toprule
\textbf{Model} & \textbf{Method} & R@10 & N@10 & Agr@10 & Agr@1 & Z@20\\
\midrule
\multirow{3}{*}{BERT} 
& Base & 0.722 & 0.531 & - & - & - \\
& Base+DFME & 0.754 & 0.514 & 0.498 & 0.220 & - \\
& GREW+DFME & \textbf{0.771} & \textbf{0.534} & \textbf{0.419} & \textbf{0.211} & \textbf{5.056}\\
\midrule
\multirow{3}{*}{SAS} 
& Base & \textbf{0.823} & \textbf{0.607} & - & - & -\\
& Base+DFME & 0.767 & 0.530 & 0.424 & 0.212 & - \\
& GREW+DFME & 0.710 & 0.447 & \textbf{0.326} & \textbf{0.233} & \textbf{6.470}\\
\bottomrule
\end{tabular}
\label{tab:mea_performance}
\end{table}

Table \ref{tab:mea_performance} (Metrics here were calculated with 100 negative samples, following the model extraction attack method, DFME \cite{yue_black-box_2021}) highlights the dual capability of GREW as both an ownership verification tool and an active defense mechanism against model extraction. We focus specifically on the Agreement metrics (Agr@10), which measure the functional similarity between the extracted model output and the victim teacher output:
\begin{equation}
\text{Agr@K} = \frac{|\mathcal{T}_{\text{top}-K} \cap \mathcal{S}_{\text{top}-K}|}{K},
\end{equation}
where $\mathcal{T}_{\text{top}-K}$ denotes the top-k recommendation list predicted by the target black-box model, and $\mathcal{S}_{\text{top}-K}$ denotes the top-k recommendation list predicted by the extracted surrogate model.

The results show a consistent decline in model extraction fidelity when the adversary targets a watermarked model compared to attacking an unwatermarked model; for instance, in the SASRec setting, Agr@10 drops significantly from 0.424 to 0.326. This indicates that the subtle perturbations introduced by the watermark effectively shift the output distribution, making it harder for the adversary to perfectly clone the teacher's recommendation ability. Crucially, despite this degradation in extraction quality, the ownership signal remains robustly embedded: the Z-scores in the extracted models reach 5.056 (BERT) and 6.470 (SAS), far exceeding the statistical significance threshold ($Z > 4.0$), ensuring that any stolen copy, however imperfect, remains undeniably traceable.

\subsection{Hyperparameter Analysis}
\begin{figure}[htbp]
  \centering
  \includegraphics[width=\linewidth]{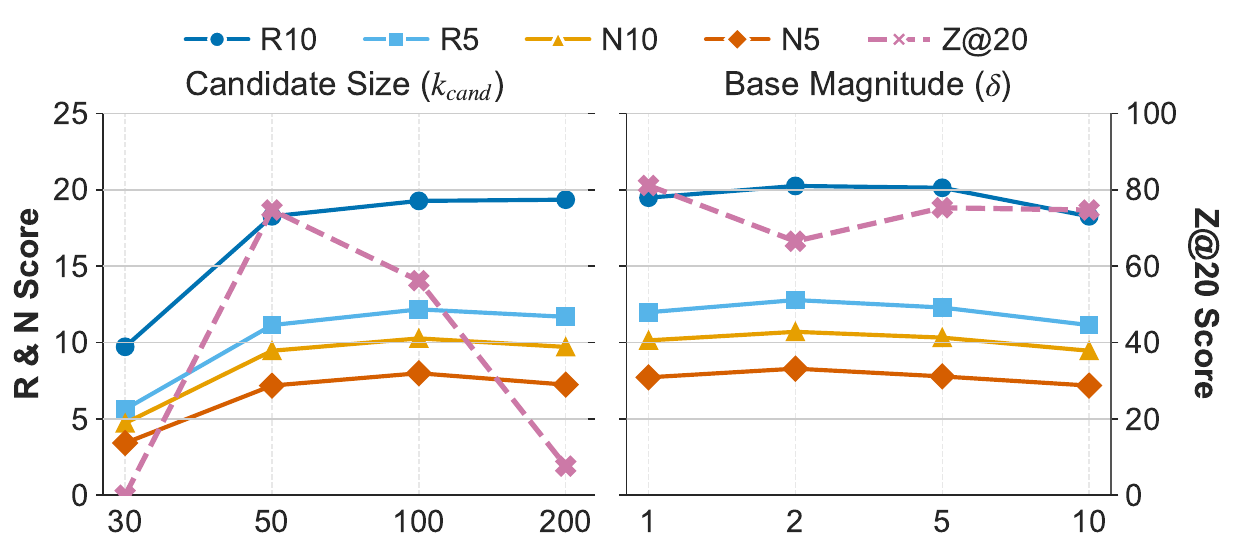}
  \caption{Watermarking performance on different hyperparameter settings under ML-1M and BERT4Rec as base model.}
  \Description{Watermarking Performance on Different Hyperparameter Settings.}
  \label{fig:parameter}
\end{figure}

We further investigate the sensitivity of GREW to two key hyperparameters: the size of the candidate item pool $k_{cand}$ for bias injection and the base watermark injection magnitude $\delta$. First, regarding the candidate pool size $k_{cand}$, we observe a fundamental trade-off between utility and verifiability. As shown in Figure \ref{fig:parameter}, an overly restrictive pool ($k_{cand}=30$) causes model collapse, reducing R@10 to 9.72\% and failing to inject a valid watermark signal. Expanding the pool from 50 to 200 restores model utility (19.35\%), but it gradually dilutes the watermark signal (yet verifiable with $P$-value $<10^{-6}$). We identify $K_{cand}=50$ as the optimal setting, achieving a high $Z$-score of 74.69 while maintaining recommendation utility. Second, regarding the base injection magnitude $\delta$, increasing $\delta$ from 1 to 10 all provides $Z$-scores high enough to claim model ownership. Notably, setting $\delta=2$ offers the most balanced performance, achieving a superior R@10 of 20.25\% while maintaining a robust verification confidence ($Z$=66.52), demonstrating that a moderate but precise bias is sufficient for effective watermarking.

\subsection{Ablation Study}
In Table \ref{tab:ablation}, we further conducted an ablation study to validate the contribution of each designed component. First, replacing the Semantic-Consistent Hash module with naive ID-based hashing ("w/o Semantic") reduces detectability (Z@20 drops from 87.39 to 68.93), confirming that, unlike simple ID-based hashing, our semantic-aligned approach enables the model to internalize the watermark as a consistent preference pattern. Second, regarding the Confidence-Adaptive Scaling mechanisms, removing the entropy-based local scaling constraint ("w/o $\alpha_{\text{local}}$") results in the lowest recommendation utility preservation (R@10 drops to 19.33\%), indicating that a uniform bias is detrimental for hard samples; similarly, the absence of global feedback control ("w/o $\alpha_{\text{global}}$") disrupts the overall equilibrium, leading to suboptimal performance in both utility and detection. Consequently, the full GREW outperforms all variants, demonstrating that semantic-aligned hashing coupled with multi-granularity (global and local) strength scaling is essential for achieving the optimal balance between verification robustness and recommendation quality.

\begin{table}[htbp]
  \caption{Watermarking performance with different variants under ML-1M and BERT4Rec as base model.}
  \label{tab:ablation}
  \begin{tabular}{cccccc}
    \toprule
    Method                & R@10 & R@5 & N@10 & N@5 & $Z$@20 \\
    \midrule
    \textbf{GREW}       & \textbf{20.45} & \textbf{12.98} & \textbf{10.68} & \textbf{8.28} & \textbf{87.39} \\
    w/o Semantic          & 19.72 & 12.30 & 10.28 & 7.90 & 68.93 \\
    w/o $\alpha_{\text{local}}$  & 19.33 & 12.34 & 10.19 & 7.96 & 74.75 \\
    w/o $\alpha_{\text{global}}$ & 19.85 & 12.33 & 10.33 & 7.92 & 85.00 \\
    \bottomrule
  \end{tabular}
\end{table}

\section{Related Work}
\subsection{Model Extraction in Recommender Systems}
Intellectual property theft in recommender systems increasingly manifests through model extraction attacks, where adversaries reconstruct proprietary algorithms by systematically exploiting public API interfaces to train high-fidelity surrogate models \cite{yue_black-box_2021}. Although deployed systems typically restrict outputs to Top-K rankings, attackers have developed sophisticated mechanisms to overcome this information bottleneck. Contemporary extraction frameworks leverage auto-regressive querying strategies and advanced generative backbones, including graph neural networks and large language models, to synthesize informative surrogate data that closely approximates the target distribution \cite{wang_sim4rec_2025, zhou_budget_2025, wang_data-free_2025}. These attacks are further amplified by optimized sampling constraints and distillation objectives, allowing adversaries to replicate complex ranking behaviors even under strict query budgets \cite{liu_fewmea_2025}. Once a high-quality surrogate is acquired, the adversaries are able to gain a surrogate white-box control, develop target attacks, or unauthorized replication. Since the memorization-based watermarks manifest as statistical outliers rather than intrinsic preferences, they are susceptible to being pruned during the surrogate learning process or explicitly removed. This vulnerability highlights the critical need for watermarking mechanisms that are intrinsically aligned with the recommendation process, ensuring persistence even when the model is subjected to high-fidelity extraction.

\subsection{Watermarking in Recommender Systems}
Model watermarking has been widely studied as a means of protecting the intellectual property of machine learning models, by embedding property-related signals or triggers into the model itself \cite{zhang2018protecting}. Existing approaches can be broadly categorized into white-box watermarking, which embeds signals into model parameters or internal representations, and black-box watermarking, which verifies ownership solely through model outputs \cite{dang_recommendation_2024}. In particular, white-box watermarking often needs modification of the model's inner weights and structures, which impractically requires full access to the model infrastructure during the detection process \cite{zhang2021deep}. On the other hand, black-box watermarking methods often rely on trigger-based inputs or predefined signals that the model is trained to memorize, with ownership verified through strong watermarked outputs \cite{wang2024defense, robinette2025trigger}. Recent progress in generative models has motivated green-red watermarking techniques that embed predefined statistical patterns into the generation process. These methods enable GenAI models to maintain normal generation quality while producing outputs that conform to owner-specified rules, allowing ownership to be verified via statistical hypothesis testing of the generated results \cite{kirchenbauer_watermark_2023, mao_watermarking_2025, lu2024entropy, panaitescu-liess_can_2025}. While these techniques have shown effectiveness in classification and generation tasks, they typically assume independent prediction targets and do not consider the structured decision mechanisms present in recommender systems.

Compared to general-purpose model watermarking, watermarking for recommender systems remains relatively underexplored. Existing studies often inject watermarks by forcing the model to memorize abnormal or rarely occurring interaction sequences \cite{zhang_watermarking_2024}, special trigger users \cite{zhang_data_2025, yang_ownership_2025}, or user-item pairs \cite{dang_recommendation_2024}. However, such designs are not naturally aligned with recommendation behaviors and may introduce noticeable distribution shifts, making the watermark susceptible to detection, removal, or performance degradation. Moreover, these approaches overlook the ranking-based nature of recommendation, treating item predictions independently rather than as competitive decisions deep in the model. These limitations motivate the need for watermarking mechanisms that are inherently compatible with recommendation decision processes.

\section{Conclusion}

We presented GREW, the first Green-Red watermarking framework designed specifically for recommender systems. Departing from the conventional strategy of injecting detectable synthetic data triggers, GREW leverages a secret key to stealthily partition the item space into green-red sets and subtly modulate the ranking probability of green items. Strong ownership verification can be carried out via key-controlled re-partitioning and statistical hypothesis testing on black-box outputs. Through the synergistic design of Semantic-Consistent Hashing, Decision-Aligned Masking, and Confidence-Adaptive Scaling, our framework successfully embeds robust watermark signals directly into the ranking behavior of a model while preserving the accuracy of Top-$K$ recommendations. Our evaluation confirms that GREW offers strong verifiability and robustness against model extraction attack. Future work may explore extending this paradigm to graph or generative recommendation architectures to further broaden the scope of intellectual property protection.


\bibliographystyle{ACM-Reference-Format}
\bibliography{library}
\balance

\appendix

\section{Dataset Statistics}
\label{appendix_a}
To provide a comprehensive overview of the data scale and structural characteristics, the detailed statistics of the datasets utilized in our empirical evaluations are summarized in Table \ref{tab:datasets}.
\begin{table}[htbp]
  \caption{Statistics of the datasets.}
  \label{tab:datasets}
  \begin{tabular}{cccccc}
    \toprule
    Dataset & \#Users & \#Items & \#Interacts & Avg.len & Density \\
    \midrule
    ML-1M   & 6,040   & 3,416   & 1,000,209   & 163.5   & 4.84\% \\
    Beauty  & 40,226  & 54,542  & 353,989     & 8.8     & 0.02\% \\
    Steam   & 334,542 & 13,046  & 3,546,145   & 10.6    & 0.10\% \\
    \bottomrule
  \end{tabular}
\end{table}

\section{Implementation Details}
The detailed training configurations for each base model are summarized in Table \ref{tab:model_config}. We employ the Adam optimizer across all architectures, with a weight decay of 0.01, a learning rate of 0.001, and a batch size of 512 (on a single NVIDIA GeForce RTX 4090 GPU with 24GB of memory). To prevent overfitting, the dropout rates for the ML-1M, Beauty, and Steam datasets are meticulously tuned to 0.1, 0.5, and 0.2, respectively. Following the protocols established in the existing methods \cite{yue_black-box_2021, zhang_watermarking_2024}, the maximum sequence lengths are set to 200 for ML-1M and 50 for both Beauty and Steam. Regarding our GREW framework, the green-item density $\gamma$ is set to 0.5, and the base injection magnitude $\delta$ is initialized at 0.1. The candidate pool size $k_{cand}$ is defined as top-100, and the global feedback scaling is configured to maintain a target aggregated hit rate $\tau=0.65$.

\begin{table}[htbp]
\centering
\caption{Model configurations. ly: layer, h: attention head, mp: masking probability.}
\begin{tabular}{ccc}
\toprule
\textbf{Phase} & \textbf{Model} & \textbf{Config. on \{ML-1M, Steam, Beauty\}} \\
\midrule
\multirow{3}{*}{Training} 
& NARM & GRU ly:1;\\
& SAS & TRM ly:2; h:2;\\
& BERT & TRM ly:2; h:2; mp=\{0.2, 0.2, 0.6\}.\\
\bottomrule
\end{tabular}
\label{tab:model_config}
\end{table}

\section{Pseudo-codes}
\label{append:alg}

\begin{algorithm}[b]
\caption{GREW Watermark Injection Procedure}
\label{alg:watermark}
\begin{algorithmic}[1]
\State \textbf{Input:} Logits $\mathbf{z}$, item embeddings $\mathbf{E}$, key $K_w$, target density $\gamma$, target hit rate $\tau$, step size $\eta$, momentum $m$.
\State \textbf{Buffers:} Global strength $\alpha_{\text{global}}$ (init: $\delta_{base}$), running hit rate $\bar{r}$.
\State \textbf{Output:} Watermarked logits $\mathbf{z}'$.

\State \textbf{// Step 1: Semantic-Consistent Hashing}
    \State $\mathbf{v}_{\text{proj}} \gets \text{PRNG}(K_w)$; \quad $\omega \gets 2\pi$
    \State $s_t \gets (a \cdot \mathbf{S}_u^{t-1} + K_w) \bmod 2^{32}, \quad \hat{s}_t \gets s_t / 2^{32}$
    \For{each item $i \in \mathcal{V}$}
        \State $c_i \gets (\mathbf{e}_i \cdot \mathbf{v}_{\text{proj}}) / \sqrt{d}$
        \State $h_i^t \gets | \sin((c_i + \hat{s}_t) \cdot \omega) |$ \Comment{Double precision mapping}
        \State $\mathbf{M}_{\text{green}, i} \gets \mathbb{I}(h_i^t < \gamma)$
    \EndFor

\State \textbf{// Step 2: Decision-Aligned Masking}
    \State $z_{\text{thresh}} \gets \text{Top-Value}(\mathbf{z}, k_{cand})$
    \State $\mathbf{M}_{\text{inject}} \gets \mathbb{I}(\mathbf{z} \ge z_{\text{thresh}}) \odot \mathbf{M}_{\text{green}}$

\State \textbf{// Step 3: Confidence-Adaptive Scaling}
    \State $p_j \gets \text{Softmax}(\mathbf{z}_{topK})$
    \State $\alpha_{\text{local}} \gets \left( -\sum p_j \log p_j / \log K \right)^{\beta}$ 
    \State $\mathbf{z}' \gets \mathbf{z} + (\alpha_{\text{global}} \cdot \alpha_{\text{local}}) \cdot \mathbf{M}_{\text{inject}}$
    \If{is\_training}
        \State $r_{curr} \gets \text{Mean}(\mathbf{M}_{\text{green}} \text{ in Top-}K \text{ of } \mathbf{z}')$
        \State $\bar{r} \gets m \cdot \bar{r} + (1 - m) \cdot r_{curr}$ \Comment{Running average update}
        \State $\alpha_{\text{global}} \gets \text{Clip}(\alpha_{\text{global}} + \eta \cdot (\tau - \bar{r}), \delta_{min}, \delta_{max})$ 
    \EndIf
\State \Return $\mathbf{z}'$
\end{algorithmic}
\end{algorithm}

\begin{algorithm}[b]
\caption{GREW Ownership Verification Procedure}
\label{alg:verification}
\begin{algorithmic}[1]
\State \textbf{Input:} Suspected model $\mathcal{M}$, test users $U$, interaction histories $\{\mathbf{S}_u^{t-1}\}_{u \in U}$, secret key $K_w$, item embeddings $\mathbf{E}$, target density $\gamma$.
\State \textbf{Output:} $Z$-score, $P$-value,

\State \textbf{Step 1: Green Set Reproduction}
\State Initialize $|s|_G \gets 0$, $N \gets |\mathcal U| \times K$
\State $\mathbf{v}_{\text{proj}} \gets \text{PRNG}(K_w)$
\For{each user $u \in U$} \Comment{Reproduce partition seed}
    \State Obtain Top-$K$ recommendation list $\mathcal{R}_u$ from $\mathcal{M}$
    \State Compute seed $\hat{s}_t$ using $K_w$ and history $\mathbf{S}_u^{t-1}$ 
    \For{each item $i \in \mathcal{R}_u$}
        \State $c_i \gets (\mathbf{e}_i \cdot \mathbf{v}_{\text{proj}}) / \sqrt{d}$
        \State $h_i^t \gets | \sin((c_i + \hat{s}_t) \cdot \omega) |$
        \If{$h_i^t \le \gamma$}
            \State $|s|_G \gets |s|_G + 1$ \Comment{Count observed green items}
        \EndIf
    \EndFor
\EndFor

\State \textbf{Step 2: Statistical Hypothesis Testing}
\State $\hat{p} \gets |s|_G / N$ \Comment{Empirical hit rate}
\State $Z \gets \frac{\hat{p} - \gamma}{\sqrt{\gamma(1 - \gamma) / N}}$ \Comment{Deviation from null hypothesis $H_0$}
\State $P \gets 1 - \Phi(Z)$ \Comment{Calculate $P$-value via standard normal CDF}

\end{algorithmic}
\end{algorithm}

\end{document}